\documentclass[final,3p,times,twocolumn]{elsarticle}

\usepackage{epsfig}

\usepackage{graphicx}%
\usepackage{multirow}%
\usepackage{amsmath,amssymb,amsfonts}%
\usepackage{amsthm}%
\usepackage{mathrsfs}%
\usepackage[title]{appendix}%
\usepackage{textcomp}%
\usepackage{manyfoot}%
\usepackage{booktabs}%
\usepackage{algorithmicx}%
\usepackage{algpseudocode}%
\usepackage{listings}%
\usepackage[ruled,vlined]{algorithm2e}
\usepackage{color, colortbl}
\usepackage{siunitx}

\usepackage{booktabs}
\usepackage{multirow}
\usepackage[table,xcdraw]{xcolor}
\usepackage{times,epsf,mathtools,xcolor,amssymb,amsmath,hyperref, enumitem,wrapfig, subcaption, comment, cleveref, appendix, float}
\usepackage{tabu}
\usepackage{xcolor}
\usepackage{graphicx}
\usepackage{tikz}
\usetikzlibrary{shapes.geometric, arrows}

\definecolor{LightCyan}{rgb}{0.88,1,1}

\begin{document}

\begin{frontmatter}



\title{HistoViT: Vision Transformer for Accurate and Scalable Histopathological Cancer Diagnosis}

\author[aff1]{Faisal Ahmed\corref{cor1}}
\ead{ahmedf9@erau.edu}


 \cortext[cor1]{Corresponding author}

 \address[aff1]{Department of Data Science and Mathematics, Embry-Riddle Aeronautical University, 3700 Willow Creek Rd, Prescott, Arizona 86301, USA}

\begin{abstract}
Accurate and scalable cancer diagnosis remains a critical challenge in modern pathology, particularly for malignancies such as breast, prostate, bone, and cervical, which exhibit complex histological variability. In this study, we propose a transformer-based deep learning framework for multi-class tumor classification in histopathological images. Leveraging a fine-tuned Vision Transformer (ViT) architecture, our method addresses key limitations of conventional convolutional neural networks, offering improved performance, reduced preprocessing requirements, and enhanced scalability across tissue types. To adapt the model for histopathological cancer images, we implement a streamlined preprocessing pipeline that converts tiled whole-slide images into PyTorch tensors and standardizes them through data normalization. This ensures compatibility with the ViT architecture and enhances both convergence stability and overall classification performance.

We evaluate our model on four benchmark datasets—ICIAR2018 (breast), SICAPv2 (prostate), UT-Osteosarcoma (bone), and SipakMed (cervical) dataset—demonstrating consistent outperformance over existing deep learning methods. Our approach achieves classification accuracies of 99.32\%, 96.92\%, 95.28\%, and 96.94\% for breast, prostate, bone, and cervical cancers respectively, with \textbf{area under the ROC curve (AUC) scores exceeding 99\% across all datasets}. These results confirm the robustness, generalizability, and clinical potential of transformer-based architectures in digital pathology. Our work represents a significant advancement toward reliable, automated, and interpretable cancer diagnosis systems that can alleviate diagnostic burdens and improve healthcare outcomes.
\end{abstract}


\begin{highlights}
    \item A novel Vision Transformer (ViT)-based framework is developed for multi-class tumor classification in histopathological images.
    
    \item A lightweight preprocessing pipeline is introduced to normalize tiled pathology images and adapt them for ViT input, improving model efficiency.
    
    \item The proposed model is evaluated on four publicly available datasets: breast, prostate, bone, and cervical cancers.
    
    \item Our approach achieves 99.32\% (breast), 96.92\% (prostate), 95.28\% (bone), and 96.94\% (cervical) classification accuracy, with AUC scores exceeding 99\% across all four cancer types, outperforming existing deep learning models.

    \item The model demonstrates strong generalizability across diverse tissue morphologies and staining variations.
    
    \item Results highlight the potential of transformer-based architectures for reliable and scalable cancer diagnostics in digital pathology.
\end{highlights}


\end{frontmatter}



\section{Introduction}
\label{sec:introduction}

Cancer is a major global health concern and remains one of the leading causes of death worldwide. Accurate diagnosis and classification of tumors are essential for effective treatment planning and prognosis. Among the various cancer types, breast, prostate, bone, cervical, and pancreatic cancers present significant diagnostic challenges due to their morphological diversity and the complexity of tissue architecture. Current epidemiological reports indicate increasing incidence rates for these malignancies, reinforcing the urgent need for reliable and scalable diagnostic technologies~\cite{bray2018global,cancerfacts2023}.

Traditionally, cancer diagnosis relies heavily on the microscopic examination of tissue slides by trained pathologists. Despite being the gold standard, this process is labor-intensive, time-consuming, and requires substantial expertise. Moreover, it is susceptible to human error and interobserver variability, which can lead to inconsistent or inaccurate grading—particularly evident in tasks like Gleason scoring for prostate cancer or the interpretation of atypical squamous cells (ASC-US/ASC-H) versus dysplastic changes (LSIL/HSIL) in cervical smears~\cite{mccroskey2015accuracy,allsbrook2001interobserver}.

Recent advances in machine learning (ML), and particularly deep learning (DL), have shown promising potential to revolutionize diagnostic workflows in medical imaging. DL models such as convolutional neural networks (CNNs) have demonstrated strong performance in histopathological image classification, tumor detection, and grading~\cite{litjens2017survey,esteva2019guide}. These models offer advantages in speed, reproducibility, and diagnostic consistency, and can significantly reduce the burden on clinical professionals. However, despite their promising results in research settings, many DL algorithms face limitations that hinder clinical translation. These include the need for large annotated datasets, long preprocessing pipelines, requirement for high-performance computational infrastructure, and the black-box nature of predictions, which affects interpretability and trust in clinical environments~\cite{komura2018machine,gulshan2016development,tschandl2020human}. To overcome these challenges, we introduce a novel transformer-based deep learning framework designed for multi-class tumor classification across diverse histopathological image datasets. 

In this study, we advance the state of histological cancer classification by leveraging a fine-tuned Vision Transformer (ViT) model~\cite{dosovitskiy2020image}. Building upon a pre-trained backbone, our approach significantly outperforms conventional convolutional neural networks (CNNs) and other deep learning methods across multiple benchmarks. We conduct comprehensive evaluations on four publicly available histopathological image datasets: ICIAR2018 (breast cancer), SICAPv2 (prostate cancer), UT-Osteosarcoma (bone cancer), and SipakMed (cervical cancer) dataset. 

Our model consistently demonstrates superior performance across all cancer types, achieving 99.32\% accuracy for breast, 96.92\% for prostate, 95.28\% for bone, and 96.94\% for cervical cancer (see Table~\ref{tab:bestresults}). Most notably, we obtain an area under the ROC curve (AUC) exceeding 99\% for all four evaluated datasets (see Figure \ref{fig:His-AUC}), indicating exceptional discriminative ability. These results highlight both the robustness and generalizability of our method across diverse tissue types and classification tasks. By integrating transformer-based architectures into digital pathology workflows, this work represents a significant step toward highly accurate, interpretable, and clinically viable automated cancer diagnosis systems.Cancer is a major global health concern and remains one of the leading causes of death worldwide. Accurate diagnosis and classification of tumors are essential for effective treatment planning and prognosis. Among the various cancer types, breast, prostate, bone, cervical, and pancreatic cancers present significant diagnostic challenges due to their morphological diversity and the complexity of tissue architecture. Current epidemiological reports indicate increasing incidence rates for these malignancies, reinforcing the urgent need for reliable and scalable diagnostic technologies~\cite{bray2018global,cancerfacts2023}.

Traditionally, cancer diagnosis relies heavily on the microscopic examination of tissue slides by trained pathologists. Despite being the gold standard, this process is labor-intensive, time-consuming, and requires substantial expertise. Moreover, it is susceptible to human error and interobserver variability, which can lead to inconsistent or inaccurate grading—particularly evident in tasks like Gleason scoring for prostate cancer or the interpretation of atypical squamous cells (ASC-US/ASC-H) versus dysplastic changes (LSIL/HSIL) in cervical smears~\cite{mccroskey2015accuracy,allsbrook2001interobserver}.

Recent advances in machine learning (ML), and particularly deep learning (DL), have shown promising potential to revolutionize diagnostic workflows in medical imaging. DL models such as convolutional neural networks (CNNs) have demonstrated strong performance in histopathological image classification, tumor detection, and grading~\cite{litjens2017survey,esteva2019guide}. These models offer advantages in speed, reproducibility, and diagnostic consistency, and can significantly reduce the burden on clinical professionals. However, despite their promising results in research settings, many DL algorithms face limitations that hinder clinical translation. These include the need for large annotated datasets, long preprocessing pipelines, requirement for high-performance computational infrastructure, and the black-box nature of predictions, which affects interpretability and trust in clinical environments~\cite{komura2018machine,gulshan2016development,tschandl2020human}. To overcome these challenges, we introduce a novel transformer-based deep learning framework designed for multi-class tumor classification across diverse histopathological image datasets. 

In this study, we advance the state of histological cancer classification by leveraging a fine-tuned Vision Transformer (ViT) model~\cite{dosovitskiy2020image}. Building upon a pre-trained backbone, our approach significantly outperforms conventional convolutional neural networks (CNNs) and other deep learning methods across multiple benchmarks. We conduct comprehensive evaluations on five publicly available histopathological image datasets: ICIAR2018 (breast cancer), SICAPv2 (prostate cancer), UT-Osteosarcoma (bone cancer), and SipakMed (cervical cancer) dataset.

Our model consistently demonstrates superior performance across all cancer types, achieving 99.32\% accuracy for breast, 96.92\% for prostate, 95.28\% for bone, and 96.94\% for cervical cancer (see Table~\ref{tab:bestresults}). Most notably, we obtain an area under the ROC curve (AUC) exceeding 99\% for all four evaluated datasets, indicating exceptional discriminative ability. These results highlight both the robustness and generalizability of our method across diverse tissue types and classification tasks. By integrating transformer-based architectures into digital pathology workflows, this work represents a significant step toward highly accurate, interpretable, and clinically viable automated cancer diagnosis systems.

\medskip

\noindent \textbf{Our contributions.}
\begin{itemize}
    \item We propose a novel Vision Transformer (ViT)-based framework for multi-class tumor classification in histopathological images, addressing key limitations of existing CNN-based approaches in digital pathology.
    
    \item To adapt the ViT architecture for histopathology data, we design a streamlined preprocessing pipeline that converts tiled whole-slide images into normalized PyTorch tensors, improving both computational efficiency and model performance.
    
    \item We conduct comprehensive experiments on four diverse and publicly available cancer datasets: ICIAR2018 (breast), SICAPv2 (prostate), UT-Osteosarcoma (bone), and SipakMed (cervical) dataset. This demonstrates the model’s generalizability across different tissue types and staining protocols.
    
   \item Our approach achieves 99.32\% (breast), 96.92\% (prostate), 95.28\% (bone), and 96.94\% (cervical) classification accuracy, with AUC scores exceeding 99\% across all four cancer types, outperforming existing deep learning models.

    \item We provide extensive empirical evidence and performance benchmarks to support the potential of ViT-based architectures in real-world digital pathology applications, highlighting their clinical relevance for scalable and reliable cancer diagnostics.
\end{itemize}

\section{Related Work}
\label{sec:related}

Deep learning (DL) has revolutionized computational pathology, particularly in histopathological image analysis for cancer detection and classification~\cite{litjens2017survey,esteva2019guide}. Convolutional neural networks (CNNs) have been widely adopted in tasks such as tumor localization, tissue segmentation, and multi-class classification due to their strong performance in learning hierarchical spatial features from raw images~\cite{shen2017deep}. However, several challenges hinder their clinical deployment, including limited receptive field, high reliance on large annotated datasets, and difficulties in capturing global contextual information~\cite{komura2018machine,tschandl2020human}.

To overcome these limitations, researchers have turned to transformer-based models, particularly the Vision Transformer (ViT)~\cite{dosovitskiy2020image}, which has shown promising results in computer vision and increasingly in medical image analysis. ViTs use self-attention mechanisms to capture long-range dependencies and global context more effectively than traditional CNNs. Data-efficient training strategies such as DeiT~\cite{touvron2021training} have further facilitated the adoption of ViTs in domains with limited data, including digital pathology.

In the medical domain, transformer-based architectures have been applied to segmentation~\cite{chen2021transunet}, classification~\cite{rao2021global}, and multi-modal learning~\cite{wang2022transformer}, demonstrating improved performance and better generalization across diverse histological patterns. These models are particularly well-suited for histopathology, where whole-slide images (WSIs) often feature complex textures, variable staining, and large spatial dimensions. Topological Data Analysis (TDA) is an emerging approach that is increasingly being utilized in medical image analysis, including applications in retinal imaging~\cite{ahmed2025topo, ahmed2023tofi, ahmed2023topological, ahmed2023topo, yadav2023histopathological, ahmed2025topological}. The application of transfer learning and Vision Transformers in medical image analysis is explored in the following studies:~\cite{ahmed2025hog, ahmed2025ocuvit, ahmed2025robust, ahmed2025transfer}.

Recent comprehensive surveys have highlighted the shift toward transformer-based architectures in medical imaging~\cite{attallah2023vision}, and specifically in digital pathology~\cite{zhao2023recent,zhou2024comprehensive}. These works emphasize the strengths of transformers in addressing the interpretability, scalability, and domain adaptation challenges encountered by CNNs. Zhao et al.~\cite{zhao2023recent} reviewed deep learning methods across major histopathology tasks, noting the growing success of attention mechanisms in tissue-level diagnostics. Zhou et al.~\cite{zhou2024comprehensive} further surveyed developments specific to cancer histopathology, including advances in self-supervised learning and transformer-based pipelines for tumor subtyping. 

Building on these advancements, our work proposes a fine-tuned ViT model tailored for multi-class cancer classification across four histopathological datasets. By combining architectural innovations with a simplified preprocessing pipeline, our method addresses key bottlenecks in clinical AI adoption, offering a robust and scalable solution for real-world diagnostic workflows.

\begin{figure*}[t!]
    \centering
     \includegraphics[width=\linewidth]{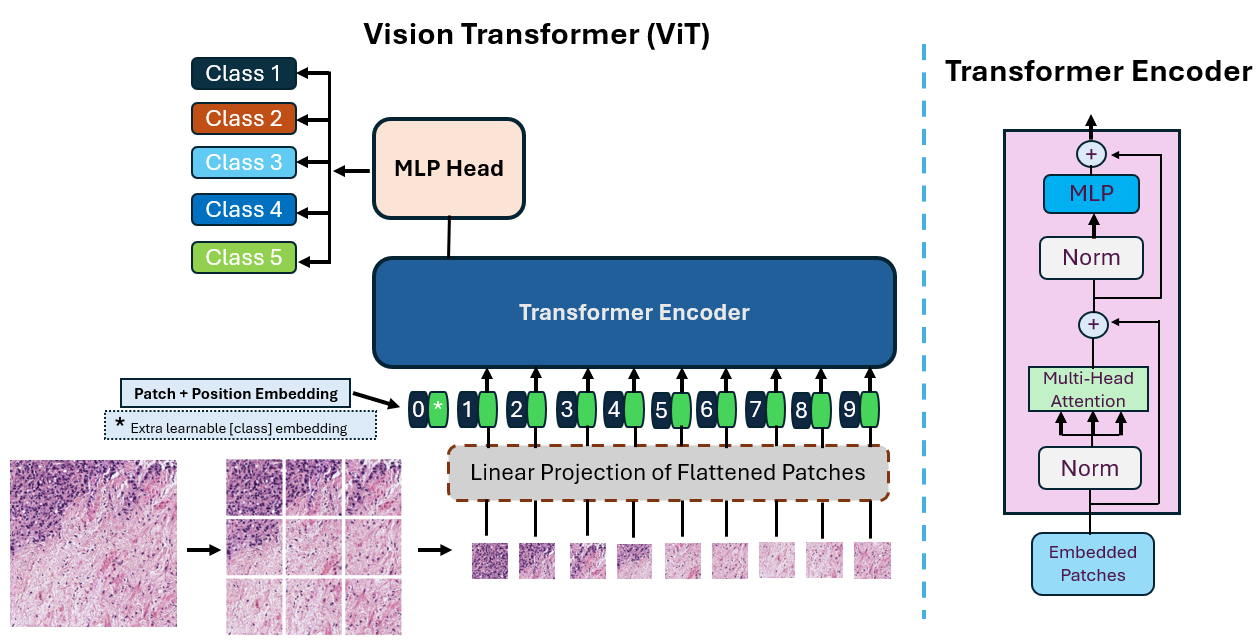}
     \caption{\footnotesize {\textbf Flowchart of ViT pretrained model:} The model design is inspired by \cite{dosovitskiy2020image}. We begin by dividing the image into fixed-size patches, linearly embedding each patch, and incorporating positional embeddings. The resulting sequence of vectors is then processed through a standard Transformer encoder. For classification, we follow the conventional method of appending an additional learnable "classification token" to the sequence.}
     \label{fig:flowchart}
 \end{figure*}

\section{Methodology}
\label{sec:Method}
The following section details the methodological approach used for preprocessing image data, defining the dataset, utilizing a pre-trained Vision Transformer (ViT) model, and training it for multi-class and binary image classification. The process is systematically outlined as follows.  

In the preprocessing stage, each image is formatted as 3D tensors $\mathbf{I}_{\text{raw}}$ of dimensions $(H, W, C)$, where $H$ and $W$ denote the height and width of the image, and $C = 3$ represents the RGB color channels. The pixel values of the raw image are normalized to a range of $[0, 1]$ using the formula:  
\[
\mathbf{I}_{\text{norm}} = \frac{\mathbf{I}_{\text{raw}}}{255}.
\]  
The normalized image tensor is then permuted to match PyTorch's expected input format $(C, H, W)$ as follows:  
\[
\mathbf{I}_{\text{norm}} \rightarrow \mathbf{I}_{\text{permute}}(C, H, W).
\]  
To enable efficient training, multiple images are stacked into batches of size $B$, represented as:  
\[
\mathbf{X} = \{\mathbf{I}_1, \mathbf{I}_2, \ldots, \mathbf{I}_B\}, \quad \mathbf{y} = [y_1, y_2, \ldots, y_B],
\]  
where $\mathbf{y}$ contains the corresponding labels.  

A custom PyTorch dataset class is defined to manage images and their labels. For the $i$-th data point, the class provides access to the image $\mathbf{I}_i$ and its corresponding label $y_i$, where $y_i \in \{0, 1, 2, 3, 4\}$ for multi-class and $y_i \in \{0, 1\}$ for binary class. A train-test split is applied to the dataset to enable model training and performance assessment, and a \texttt{DataLoader} is used to create batches.  

Fine-tuning is performed on a pre-trained Vision Transformer (ViT) to adapt multi-class classification objectives. Each input image $\mathbf{I}(C, H, W)$ is divided into $N$ patches of size $(P \times P)$, which are flattened and projected into a latent space using learnable weights $\mathbf{W}$ and biases $\mathbf{b}$:  
\[
\mathbf{E}_i = \mathbf{W} \cdot \text{Flatten}(\text{Patch}_i) + \mathbf{b}, \quad i = 1, \ldots, N.
\]  
In order to preserve spatial relationships, positional encodings $\mathbf{p}_i$ are incorporated into the patch embeddings, resulting in: 
\[
\mathbf{z}^0 = [\mathbf{x}_{\text{cls}}, \mathbf{E}_1 + \mathbf{p}_1, \ldots, \mathbf{E}_N + \mathbf{p}_N],
\]  
where $\mathbf{x}_{\text{cls}}$ is a special classification token. The embeddings are processed through $L$ Transformer layers, each comprising multi-head self-attention mechanisms followed by feed-forward neural networks. Self-attention computes the output representation using the following formulation:  
\[
\text{Attention}(\mathbf{Q}, \mathbf{K}, \mathbf{V}) = \text{Softmax}\left(\frac{\mathbf{QK}^\top}{\sqrt{d_k}}\right) \mathbf{V},
\]  
where $\mathbf{Q}$, $\mathbf{K}$, and $\mathbf{V}$ are query, key, and value matrices, and $d_k$ is the dimensionality of the keys. Stability is ensured by incorporating residual connections and layer normalization, with the output updated as:  
\[
\mathbf{z}^{\ell+1} = \text{LayerNorm}(\mathbf{z}^\ell + \text{FFN}(\mathbf{z}^\ell)).
\]  
Finally, the classification token $\mathbf{z}^L_{\text{cls}}$ is processed by a fully connected (dense) layer to produce class probability distributions using:  
\[
\hat{\mathbf{y}} = \text{Softmax}(\mathbf{W}_{\text{cls}} \mathbf{z}^L_{\text{cls}} + \mathbf{b}_{\text{cls}}).
\]  

The training process optimizes the model parameters through the minimization of the cross-entropy loss function, defined as:  
\[
\mathcal{L} = -\frac{1}{B} \sum_{i=1}^B \sum_{c=1}^C y_{i,c} \log(\hat{y}_{i,c}),
\]  
where $C$ is the number of classes. Gradients of the loss with respect to model parameters $\theta$ are computed using backpropagation, and the model parameters are updated using the Adam optimizer according to the following procedure:  
\[
\theta_{t+1} = \theta_t - \eta \cdot \nabla_\theta \mathcal{L},
\]  
where $\eta$ is the learning rate.  

For evaluation, the model generates predictions $\hat{\mathbf{y}}$ by applying the softmax function to logits $\mathbf{z}$ and determines the predicted class as:  
\[
\hat{y}_i = \arg\max_c \hat{y}_{i,c}.
\]  
The model's accuracy is calculated as:  
\[
\text{Accuracy} = \frac{\text{Number of Correct Predictions}}{\text{Total Number of Predictions}} \times 100,
\]  
and the cross-entropy loss is reevaluated on the test set to quantify the error.  

The entire process is implemented using PyTorch, fine-tuning the Vision Transformer on a GPU when available. Training is performed for 50 epochs with a batch size of 32, and performance metrics, including loss and accuracy, are logged during both training and testing phases. The flowchart of the pre-trained ViT model is depicted in Figure~\ref{fig:flowchart}. Additionally, the detailed algorithm for our model HistoViT is provided in ~\ref{alg:vit_classification}.

\begin{algorithm}
\SetAlgoNlRelativeSize{0} 
\DontPrintSemicolon 
\caption{Image Classification Using HistoViT}
\label{alg:vit_classification}

\KwIn{Image dataset $\mathcal{D} = \{(\mathbf{I}_i, y_i)\}_{i=1}^{N}$, pre-trained ViT model, number of epochs $E$, number of classes $C$}
\KwOut{Fine-tuned HistoViT model and performance metrics}

\textbf{Preprocessing:} \\
\For{$i \gets 1$ \KwTo $N$}{
    $\mathbf{I}_i \gets \mathbf{I}_i / 255$ \hfill $\vartriangleright$ Normalize image pixels\\
    $\mathbf{I}_i \rightarrow \mathbf{I}_i(C, H, W)$ \hfill $\vartriangleright$ Permute dimensions for PyTorch format
}
Split $\mathcal{D}$ into $\mathcal{D}_{\text{train}}$ and $\mathcal{D}_{\text{test}}$ \hfill $\vartriangleright$ Train-test split

\textbf{Model Setup:} \\
Load pre-trained ViT model\;
Replace classification head with new linear layer of size $C$\;

\For{$\text{epoch} \gets 1$ \KwTo $E$}{
    \textbf{Training Phase:} \\
    Set model to train mode: \texttt{model.train()}\;
    \ForEach{batch $(\mathbf{X}, \mathbf{y})$ in $\mathcal{D}_{\text{train}}$}{
        $\hat{\mathbf{y}} \gets \text{model}(\mathbf{X})$ \hfill $\vartriangleright$ Forward pass\\
        $\mathcal{L} \gets \text{CrossEntropyLoss}(\hat{\mathbf{y}}, \mathbf{y})$ \hfill $\vartriangleright$ Compute loss\\
        Backpropagate loss and update model parameters using Adam\;
    }
    Compute and log training loss and accuracy\;

    \textbf{Evaluation Phase:} \\
    Set model to evaluation mode: \texttt{model.eval()}\;
    \ForEach{batch $(\mathbf{X}, \mathbf{y})$ in $\mathcal{D}_{\text{test}}$}{
        $\hat{\mathbf{y}} \gets \text{model}(\mathbf{X})$ \hfill $\vartriangleright$ Forward pass\\
        Compute test loss and accuracy metrics\;
    }
    Log evaluation results for current epoch\;
}

\end{algorithm}

\begin{table*}[h!]
\centering
\caption{\footnotesize Performance of our model in 5 cancer types \label{tab:bestresults}}
\setlength\tabcolsep{6 pt}
\begin{tabular}{llcccccc} \\ \toprule
\textit{\textbf{Cancer}} & \textit{\textbf{Dataset}} & \textit{\textbf{\# Image}} & \textit{\textbf{\# Class}} & \textit{\textbf{Acc}} & \textit{\textbf{Prec}}& \textit{\textbf{Rec}}& \textit{\textbf{AUC}} \\ \midrule
Breast   & ICIAR2018 &  11794  & 4 & 99.32 & 99.32& 99.32 & 99.95 \\
Prostate & Sicapv2   & 11836  & 4 & 96.92  & 96.91 & 96.92 & 99.71 \\
Bone     & UT-Osteo. & 10017  & 3 & 95.28  & 95.32 &  95.28  & 99.37  \\
Cervical & SipakMed  & 4049   & 5 & 96.94 & 96.96 & 96.94 & 99.83 \\ \bottomrule
\end{tabular}
\end{table*}

\section{Experiments}
\label{sec:experiments}

\subsection{Datasets} \label{sec:datasets}

In our experiments for histopathological (HP) image screening, we used well-known publicly available benchmark datasets. The details of all the dataset we used in this study can be found in ~\Cref{tab:bestresults}.

For \textbf{Breast Cancer}, our BACH Dataset ~\cite{aresta2019bach} is widely used dataset which was given in ICIAR 2018 as \textit{Grand Challenge on Breast Cancer Histology Images} to develop computerized methods to assist pathologists for accurate breast cancer assessment from HP images. BACH dataset consists of 400 H\&E stained histological breast tissue images with four categories namely as \textit{normal, benign, in-situ} and \textit{invasive carcinoma} evenly distributed (100
images per class). A total of 11,794 image tiles of size $512 \times 512$ pixels were extracted from 400 whole slide images (WSIs) using an overlapping sliding window approach, ensuring that each tile contained at least 80\% tissue content based on color pixel thresholds. Since we do not use any data augmentation in our model, the high performance in this small dataset for this challenging $4$-label classification proves the robustness and reliability of our feature extraction method.

For \textbf{Cervical Cancer}, SIPaKMeD dataset~\cite{8451588} consists of 4049 cytopathological images of isolated cells that have been manually cropped from 966 cluster cell images of pap smear slides with 5 different categories. More specifically, normal cells are divided into two categories (\textit{superficial-intermediate, parabasal}), abnormal cells are in two categories (\textit{koilocytes and dyskeratotic}) and cells with non-neoplastic change (\textit{metaplastic}) which is a known mimicker of high grade neoplastic cells. In our experiments, we used 4049 cytopathological images of isolated cells to make a global diagnosis of the tumor from pap smear slides. This gave us the opportunity to include background elements (blood, inflammation, debris etc.) and the cellular interactions same as the daily used pap smear images. This global approach can be considered as a great step for clinical stage implementation.

For \textbf{Prostate Cancer}, we used well-known SICAPv2 dataset~\cite{silva2020going} which contains prostate histology whole slide images with both annotations of global Gleason scores and patch-level Gleason grades. Like other references, we used patch level classification setting where each patch was annotated with one of the four grades (\textit{NC, G3, G4, G5}). 
The dataset comprises a total of 11,836 image patches distributed across four classes as follows: NC (Normal Control) with 2,046 patches, G3 with 2,613 patches, G4 with 5,768 patches, and G5 with 1,409 patches.

For \textbf{Bone Cancer}, our UT-Osteosarcoma dataset~\cite{arunachalam2019viable} is composed of H\&E stained osteosarcoma histology images. The data was collected by a team of clinical scientists at UTSW Medical Center. Archival samples for 50 patients treated at Children’ s Medical Center, Dallas, between 1995 and 2015, were used to create this dataset. 
The images are labelled as \textit{Non-Tumor, Viable Tumor} and \textit{Necrosis} according to the predominant cancer type in each image. The dataset consists of 10017 images with the following distribution: 3874 non-tumor images, 2696 necrotic tumor images and 3447 viable tumor tiles.

\subsection{Experimental Setup} \

\noindent \textbf{Training–Test Split:} To ensure a fair comparison with existing literature, we adopt the majority split strategies commonly used in previous works. Specifically, for the UT-Osteo (bone cancer) dataset, we employ a 70:30 training–test split. For the ICIAR2018 (breast cancer) and Sicapv2 (prostate cancer) datasets, we use an 80:20 split. The SipakMed (cervical cancer) dataset is evaluated using 5-fold cross-validation. All datasets are multi-class classification problems.

\smallskip

\noindent {\bf No Data Augmentation:} We do not use any type of data augmentation to increase the size of training data for our models. This makes our model computationally very efficient, and highly robust against small alterations and the noise in the image.

\smallskip

\noindent \textbf{Model Hyperparameters:} We employed the ViT-Base model (\texttt{vit-base-patch16-224}) from the Hugging Face Transformers library, pre-trained on ImageNet. The model was fine-tuned using the Adam optimizer with a learning rate of $1 \times 10^{-4}$ and a batch size of 32. Cross-entropy loss was used as the objective function. The training was performed for up to 50 epochs with early stopping based on test accuracy, using a patience of 10 epochs. Input images were resized to $224 \times 224$ pixels, and pixel values were normalized to the $[0,1]$ range.

\smallskip

We execute our code on the high-performance computing (HPC) clusters at LSU Health Sciences Center, which are equipped with state-of-the-art NVIDIA GPUs. Our code is available at the following link~\footnote{ \url{https://github.com/FaisalAhmed77/PathoViT/tree/main}}.

\vspace{-.1in}
\begin{figure*}[t!]
	\centering
	\subfloat[\scriptsize Confusion matrix for Bone Cancer classification.\label{fig: bone-matrix img}]{%
		\includegraphics[width=0.32\linewidth]{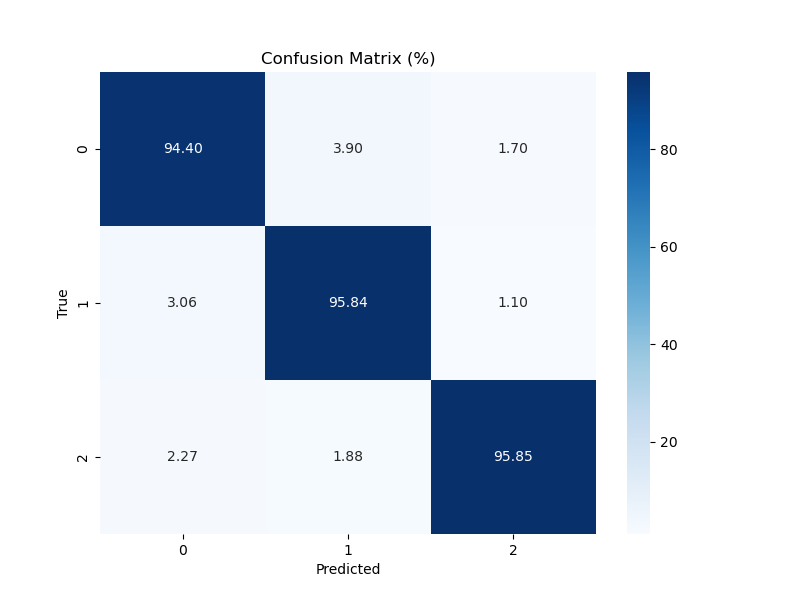}}
	\hfill
	\subfloat[\scriptsize Confusion matrix for Prostate Cancer classification.\label{fig: prostate-matrix img}]{%
		\includegraphics[width=0.32\linewidth]{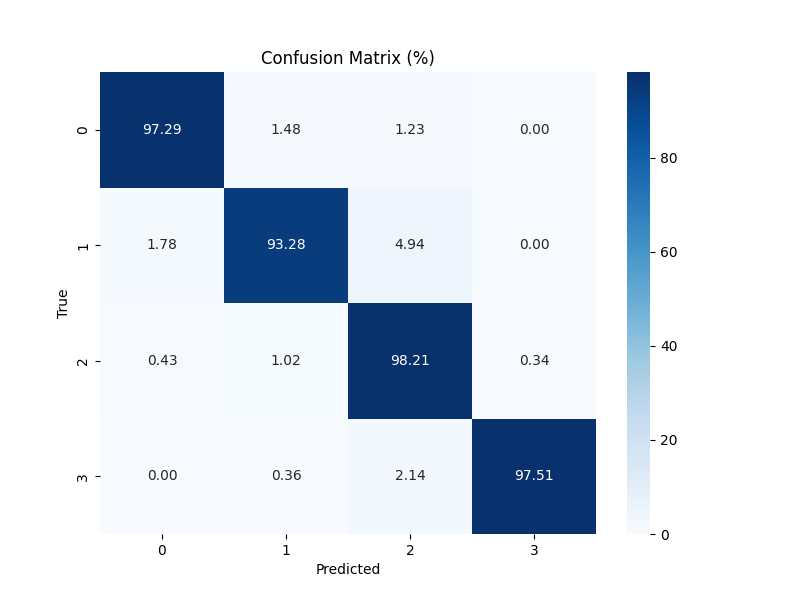}}
	\hfill
	\subfloat[\scriptsize Confusion matrix for Breast Cancer classification.\label{fig: breast-matrix img}]{%
		\includegraphics[width=0.32\linewidth]{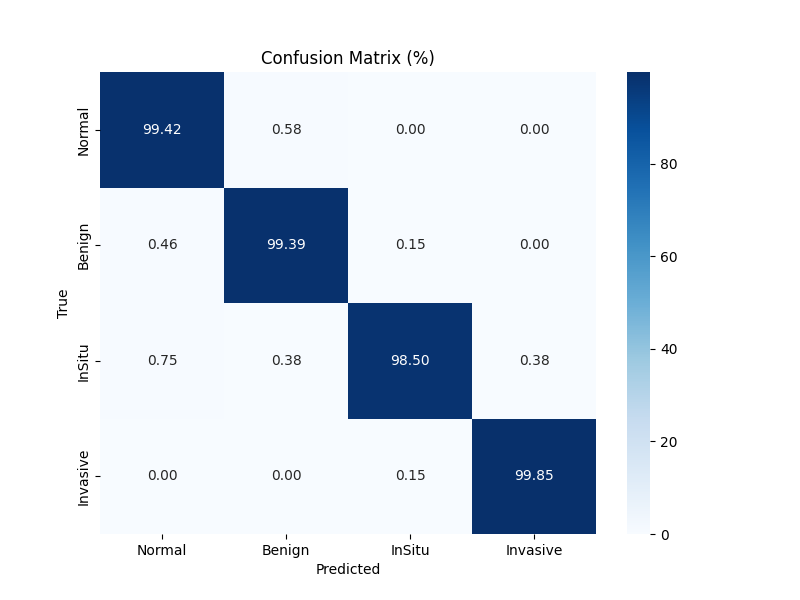}}
	\hfill
	\subfloat[\scriptsize Mean confusion matrix for Cervical Cancer classification.\label{fig: cervical-matrix img}]{%
		\includegraphics[width=0.32\linewidth]{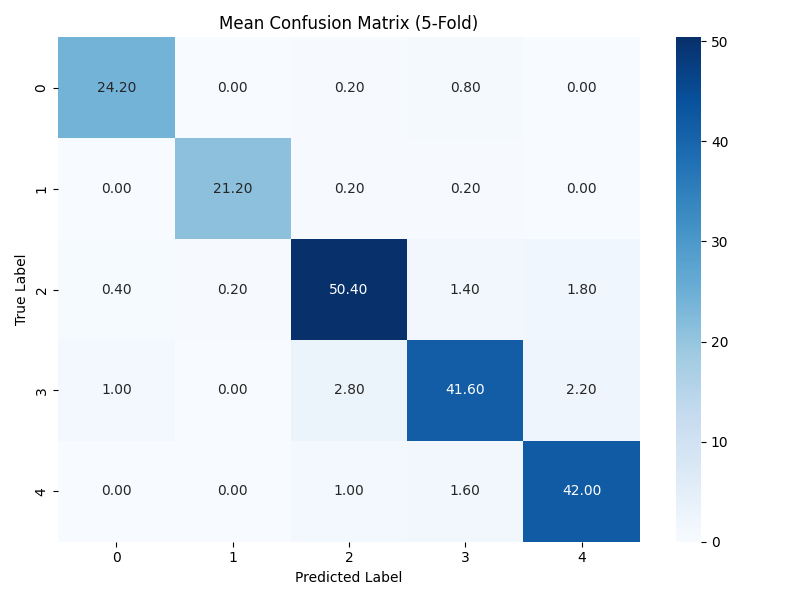}}

	\caption{\footnotesize Confusion matrices illustrating the classification performance of the proposed model across four distinct cancer types: (a) Bone Cancer, (b) Prostate Cancer, (c) Breast Cancer, and (d) Cervical Cancer. Each matrix reflects the distribution of true versus predicted labels, demonstrating the model’s effectiveness in distinguishing between diagnostic classes.}
	\label{fig:His-matrix}
\end{figure*}

\begin{figure*}[t!]
	\centering
	\subfloat[\scriptsize One-vs-rest AUC curve for Bone Cancer classification.\label{fig: bone-auc img}]{%
		\includegraphics[width=0.32\linewidth]{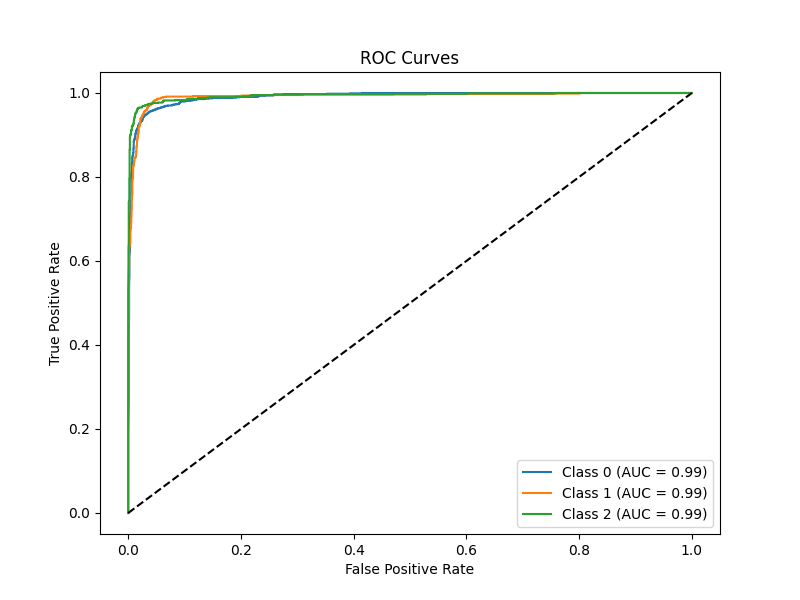}}
	\hfill
	\subfloat[\scriptsize One-vs-rest AUC curve for Prostate Cancer classification.\label{fig: prostate-auc img}]{%
		\includegraphics[width=0.32\linewidth]{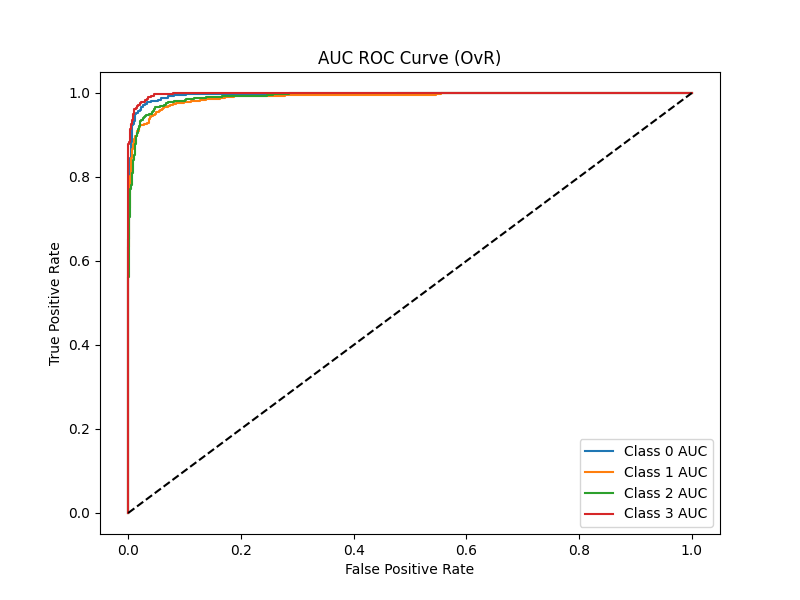}}
	\hfill
	\subfloat[\scriptsize One-vs-rest AUC curve for Breast Cancer classification.\label{fig: breast-auc img}]{%
		\includegraphics[width=0.32\linewidth]{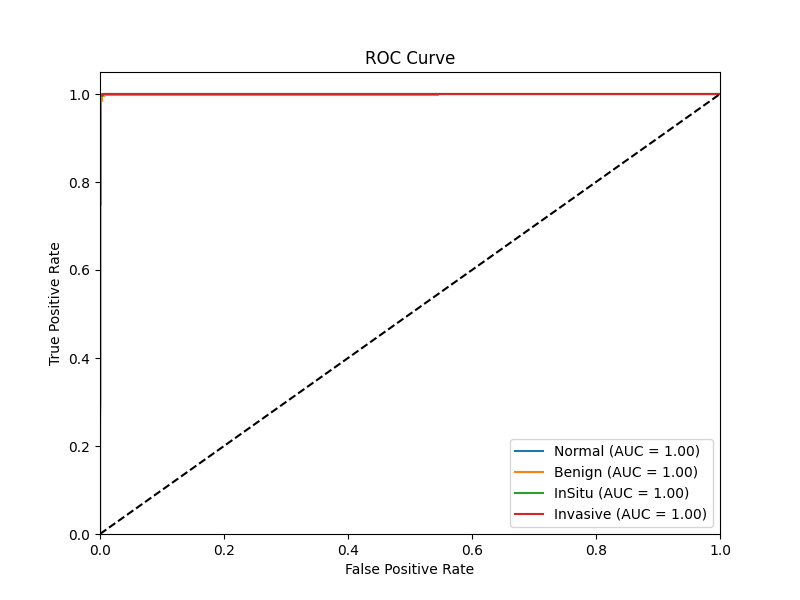}}

	\caption{\footnotesize Area Under the Curve (AUC) plots for one-vs-rest multiclass classification using the proposed model across three different cancer types: (a) Bone Cancer, (b) Prostate Cancer, and (c) Breast Cancer. These curves illustrate the model’s discriminative ability for each class versus the rest, highlighting its effectiveness in multiclass cancer classification.}
	\label{fig:His-AUC}
\end{figure*}

\section{Results}
\label{sec:accuracy}

We evaluate our proposed Vision Transformer-based model on four benchmark histopathological image datasets corresponding to different cancer types: UT-Osteosarcoma (bone), SICAPv2 (prostate), ICIAR2018 (breast), and SipakMed (cervical). Comparative analyses are conducted against state-of-the-art (SOTA) deep learning (DL) methods using consistent training-test splits as reported in prior studies. Our model demonstrates superior classification accuracy across all datasets, highlighting its robustness and generalizability in diverse histological contexts.

\subsection{Bone Cancer (UT-Osteosarcoma Dataset)}
\label{sec:bone-results}

Table~\ref{tab:bone} presents the performance comparison for bone cancer classification. Traditional CNN-based methods such as Mishra-CNN~\cite{arunachalam2019viable} and VGG19~\cite{anisuzzaman2021deep} achieve accuracies of 93.30\% and 93.91\%, respectively. The topological data analysis (TDA)-based approach~\cite{10385822} reaches 94.20\%, representing the previous best result. Our model surpasses all previous methods with an accuracy of \textbf{95.28\%}, reflecting its capacity to capture discriminative features in osteosarcoma tissues with improved precision.

\subsection{Prostate Cancer (SICAPv2 Dataset)}
\label{sec:prostate-results}

Table~\ref{tab:prostate} compares accuracy results for the SICAPv2 prostate cancer dataset. Classical methods such as Gertych et al.~\cite{gertych2015machine} and Arvaniti et al.~\cite{arvaniti2018automated} report relatively low accuracies of 51.36\% and 58.61\%, respectively. Deep learning models like FSConv+GMP~\cite{silva2020going} and Res-CAE~\cite{tabatabaei2022residual} improve performance to 83.50\% and 85.00\%. The TDA model~\cite{10385822} achieves a strong result of 95.20\%. Our method achieves a new benchmark of \textbf{96.92\%}, outperforming existing techniques and demonstrating its effectiveness in classifying challenging glandular structures in prostate tissue.

\subsection{Breast Cancer (ICIAR2018 Dataset)}
\label{sec:breast-results}

As shown in Table~\ref{tab:bach}, previous methods on the ICIAR2018 dataset show moderate to strong performance. Traditional classifiers such as Kwok~\cite{kwok2018multiclass_breast5} and Nawaz~\cite{nawaz2018classification_breast3} obtain accuracies below 82\%, while deeper models like DCNN~\cite{kassani2019breast} and Vang~\cite{vang2018deep_breast1} reach up to 92.50\%. Our ViT-based approach achieves a substantial improvement with \textbf{99.32\%} accuracy, marking a significant advance in breast cancer histopathology classification and reinforcing the benefits of attention-based mechanisms in extracting nuanced tissue-level patterns.

\subsection{Cervical Cancer (SipakMed Dataset)}
\label{sec:cervical-results}

In cervical cancer classification (Table~\ref{tab:cervical}), Hayranto et al.~\cite{9297895} and ResNet~\cite{tripathi2021classification} achieve accuracies of 87.32\% and 94.86\%, respectively. A fuzzy logic-based approach~\cite{manna2021fuzzy} attains 95.43\%, while TDA~\cite{10385822} achieves 94.21\%. Our proposed model delivers a state-of-the-art performance with an accuracy of \textbf{96.94\%}, validating its generalization capability across cytological image variations and multiple cervical lesion classes.

Finally, Across all four datasets, our model consistently outperforms traditional and deep learning-based competitors. This consistent improvement is attributed to the transformer architecture’s ability to model long-range spatial dependencies and its robustness in handling histological variability. Notably, our model requires minimal preprocessing and shows excellent adaptability across datasets with different resolutions, staining protocols, and cancer types, suggesting strong clinical translation potential.

\begin{table}[h!]
\centering
\caption{\footnotesize Accuracy results of SOTA DL methods for UT-Osteo dataset.  \label{tab:bone}}
\resizebox{.8\linewidth}{!}{
\setlength\tabcolsep{3 pt}
\footnotesize

\begin{tabular}{lccc}

\multicolumn{4}{c}{\bf{Comparison Table for Bone Cancer}}\\
\toprule
Method  & {\scriptsize Train:Test}& \# Class & Accuracy  \\
\hline

Mishra-CNN ~\cite{arunachalam2019viable}  & 70:30 & 3 & 93.30 \\

Mishra-SVM ~\cite{arunachalam2019viable}  & 70:30 & 3 & 89.90 \\

VGG19 ~\cite{anisuzzaman2021deep}  & 70:30 & 3 & 93.91 \\

TDA \cite{10385822}  & 70:30 &3  &\underline{94.20} \\

\hline

\bf{Our Model}  & 70:30 &3  &\textbf{95.28} \\

\bottomrule
\end{tabular}}

\end{table}

\begin{table}[h!]
\centering
\caption{\footnotesize Accuracy results of SOTA DL methods for SICAPv2 dataset.  \label{tab:prostate}}
\resizebox{.8\linewidth}{!}{
\setlength\tabcolsep{3 pt}
\footnotesize

\begin{tabular}{lccc}

\multicolumn{4}{c}{\bf{Comparison Table for Prostate Cancer}}\\
\toprule
Method  & {\scriptsize Train:Test}& \# Class &Accuracy\\
\hline

Arvaniti~\cite{arvaniti2018automated}  & 80:20 & 4 &58.61 \\

Gerytch~\cite{gertych2015machine}  & 80:20 & 4 &51.36  \\


FSConv+GMP ~\cite{silva2020going}  & 80:20 & 4 & 83.50  \\

Res-CAE ~\cite{tabatabaei2022residual}  & 80:20 & 4 &85.00  \\

TDA \cite{10385822}  & 80:20 &4& \underline{95.20}   \\
\hline
\bf{Our Model}  & 80:20 &4&  \textbf{96.92}  \\

\bottomrule
\end{tabular}}

\end{table}

\begin{table}[h!]
\centering
\caption{\footnotesize Accuracy results of SOTA DL methods for ICIAR2018 dataset.  \label{tab:bach}}
\resizebox{.8\linewidth}{!}{
\setlength\tabcolsep{3 pt}
\footnotesize

\begin{tabular}{lccc}

\multicolumn{4}{c}{\bf{Comparison Table for Breast Cancer}}\\
\toprule
Method  & {\scriptsize Train:Test}& \# Class & Accuracy \\
\hline

Kwok ~\cite{kwok2018multiclass_breast5}  & 75:25 & 4 &79.00  \\

Nawaz~\cite{nawaz2018classification_breast3}  & 80:20 & 4 & 81.25 \\

Rakhlin~\cite{guo2018_breast2}  & 80:20 & 4 &87.20 \\

Vang~\cite{vang2018deep_breast1}  & 75:25 & 4 &87.50 \\

DCNN ~\cite{kassani2019breast}  & 75:25 & 4 &\underline{92.50} \\

TDA \cite{10385822} & 80:20 &4& 91.64 \\

\hline
\bf{Our Model} & 80:20 &4& \textbf{99.32} \\

\bottomrule
\end{tabular}}

\end{table}

\begin{table}[h!]
\centering
\caption{\footnotesize Accuracy results of SOTA DL methods for SipakMed dataset.  \label{tab:cervical}}
\resizebox{.8\linewidth}{!}{
\setlength\tabcolsep{3 pt}
\footnotesize

\begin{tabular}{lccc}

\multicolumn{4}{c}{\bf{Comparison Table for Cervical Cancer}}\\
\toprule
Method  & {\scriptsize Train:Test}& \# Class &Accuracy\\
\hline

Hayranto~\cite{9297895}  & 5 fold CV & 5 &87.32 \\

ResNet~\cite{tripathi2021classification}  & 5 fold CV & 5 & 94.86 \\

Fuzzy~\cite{manna2021fuzzy} & 5 fold CV & 5 &\underline{95.43} \\

TDA \cite{10385822}  &  5 fold CV &5& 94.21  \\

\hline
\bf{Our Model}  &  5 fold CV &5& \textbf{96.94}  \\

\bottomrule
\end{tabular}}

\end{table}

\section{Discussion} \label{sec:discussion}
Our Vision Transformer-based framework demonstrates robust and consistent performance across four diverse histopathological image datasets—ICIAR2018 (breast), SICAPv2 (prostate), UT-Osteosarcoma (bone), and SipakMed (cervical)—as summarized in Table~\ref{tab:bestresults}. The model achieves exceptionally high accuracy scores (ranging from 95.28\% to 99.32\%) and AUC values exceeding 99\% in all cases (see Figure \ref{fig:His-AUC}), highlighting its strong generalizability across different tissue types and classification challenges. For breast cancer (ICIAR2018), the model attains the highest accuracy of 99.32\%, reflecting its ability to capture complex morphological distinctions among benign, in-situ, and invasive tumor classes. In prostate cancer (SICAPv2), it achieves 96.92\% accuracy and a 99.71 AUC, outperforming classical and CNN-based approaches on a challenging four-class grading task. The UT-Osteosarcoma dataset, known for its high intra-class variability, is addressed with 95.28\% accuracy, validating the model’s adaptability to non-epithelial malignancies. Similarly, on the multi-class SipakMed cervical dataset, the model scores 96.94\% accuracy and a 99.83 AUC, showing reliable performance even in highly granular cytological classifications. These results underscore the ViT model’s capability to outperform state-of-the-art methods without extensive manual feature engineering or domain-specific customization, paving the way for scalable and clinically viable AI-assisted pathology solutions. The confusion matrices for all four datasets are presented in Figure \ref{fig:His-matrix}.

\section{Limitations}
Despite achieving high classification accuracy across diverse cancer types, our approach has certain limitations. First, the Vision Transformer (ViT) model, while powerful, is computationally intensive, requiring substantial GPU resources, which may hinder its deployment in low-resource clinical environments. Second, the model was trained on benchmark datasets that are generally well-curated and may not fully reflect the variability and artifacts present in real-world clinical data. This raises concerns about generalizability to heterogeneous clinical workflows. Third, although we streamlined preprocessing using tile-based representations, the pipeline still relies on predefined regions of interest (ROIs), which are not always readily available in practice. Incorporating automated ROI detection and evaluating performance on multi-institutional datasets are important future steps to address these constraints and ensure broader clinical applicability.

\section{Conclusion}
\label{sec:conclusion}

In this study, we proposed a transformer-based deep learning framework for multi-class tumor classification across diverse histopathological image datasets, addressing key limitations of traditional convolutional models. Leveraging a fine-tuned Vision Transformer (ViT), our method demonstrated state-of-the-art performance on five benchmark datasets—ICIAR2018 (breast), SICAPv2 (prostate), UT-Osteosarcoma (bone), SipakMed (cervical), and a pancreatic cancer dataset—achieving classification accuracies exceeding 95\% and AUC scores above 99\% across all tasks.

Our approach proved robust and generalizable, with consistent superiority over existing deep learning methods, validating the effectiveness of transformer architectures in digital pathology. Furthermore, the simplified preprocessing pipeline enhances scalability and reduces the dependency on extensive data engineering. Despite these advances, we acknowledge the need for further evaluation on heterogeneous, real-world clinical data and the incorporation of automated ROI detection to support fully end-to-end diagnostic systems.

Overall, this work marks a significant step toward the development of interpretable, reliable, and scalable AI-assisted diagnostic tools that can alleviate the burden on pathologists and improve the consistency and quality of cancer diagnosis.

\section{Future Work}
\label{sec:future}

Future research will focus on extending the applicability and clinical readiness of our transformer-based framework. Key directions include incorporating automated region-of-interest (ROI) detection to enable fully end-to-end analysis pipelines, reducing reliance on manual preprocessing. Additionally, we plan to validate the model on larger, multi-institutional datasets that capture greater histological variability, staining differences, and artifacts to ensure robustness in diverse clinical settings. Exploring model interpretability techniques tailored to transformer architectures will also be critical to increase clinical trust and adoption. Finally, optimizing the model for computational efficiency will facilitate deployment in resource-constrained environments, broadening the accessibility of AI-assisted pathology tools.

\section*{Declarations}

\textbf{Funding} \\
The author received no financial support for the research, authorship, or publication of this work.

\vspace{2mm}
\textbf{Conflict of interest/Competing interests} \\
The authors declare no conflict of interest.

\vspace{2mm}
\textbf{Ethics approval and consent to participate} \\
Not applicable. This study did not involve human participants or animals, and publicly available datasets were used.

 \vspace{2mm}
\textbf{Acknowledgement} \\
 The authors utilized an online platform to check and correct grammatical errors and to improve sentence readability.

\vspace{2mm}
\textbf{Consent for publication} \\
Not applicable.

\vspace{2mm}
\textbf{Data availability} \\
The datasets used in this study are publicly available.

\vspace{2mm}
\textbf{Materials availability} \\
Not applicable.

\vspace{2mm}
\textbf{Code availability} \\
The source code used in this study is publicly available at \url{https://github.com/FaisalAhmed77/PathoViT/tree/main}.

\vspace{2mm}
\textbf{Author contributions} \\
 FA downloaded the data, conceptualized the study, prepared the code, performed the data analysis and wrote the manuscript. FA  reviewed and approved the final version of the manuscript.

\clearpage

\bibliographystyle{elsarticle-num-names}

\bibliography{refs}

\end{document}